\documentclass[aps,prb,showpacs,amsmath,amssymb,twocolumn,notitlepage]{revtex4-2}

\usepackage{color}
\usepackage{amsmath}
\usepackage{graphicx}
\usepackage{multirow}
\usepackage{float}
\usepackage{bm}

\begin{document}

\newcommand{\bn}{{\bm n}}
\newcommand{\bp}{{\bm p}}   
\newcommand{\br}{{\bm r}}
\newcommand{\bk}{{\bm k}}
\newcommand{\bv}{{\bm v}}
\newcommand{\brho}{{\bm{\rho}}}
\newcommand{\bj}{{\bm j}}
\newcommand{\wk}{\omega_{\bf k}}
\newcommand{\nk}{n_{\bf k}}
\newcommand{\eps}{\varepsilon}
\newcommand{\la}{\langle}
\newcommand{\ra}{\rangle}
\newcommand{\be}{\begin{equation}}
\newcommand{\ee}{\end{equation}}
\newcommand{\intl}{\int\limits_{-\infty}^{\infty}}
\newcommand{\dE}{\delta{\cal E}^{ext}}
\newcommand{\SE}{S_{\cal E}^{ext}}
\newcommand{\dsp}{\displaystyle}
\newcommand{\phit}{\varphi_{\tau}}
\newcommand{\p}{\varphi}
\newcommand{\cL}{{\cal L}}
\newcommand{\dphi}{\delta\varphi}
\newcommand{\dbj}{\delta{\bf j}}
\newcommand{\dI}{\delta I}
\newcommand{\dph}{\delta\varphi}
\newcommand{\ua}{\uparrow}
\newcommand{\da}{\downarrow}
\newcommand{\ip}{\{i_{+}\}}
\newcommand{\im}{\{i_{-}\}}
\newcommand{\lambdas}{n\ldots n_3}
\newcommand{\nnn}[1]{{\color{blue} #1}}
\renewcommand{\nnn}[1]{ #1}
  
\title{Electron--electron scattering and conductivity of disordered systems with Galilean-invariant spectrum}

\author{K. E.~Nagaev} 

\affiliation{Kotelnikov Institute of Radioengineering and Electronics, Mokhovaya 11-7, Moscow 125009, Russia}
\affiliation{National Research University Higher School of Economics, Moscow 101978, Russia}

\date{\today}

\begin{abstract}
The electron--electron scattering does not affect the electrical current in Galilean--invariant systems.
We show that nevertheless electron--electron collisions may contribute to the electric resistivity of systems with 
parabolic spectrum
provided that they have multiply connected Fermi surface and there is an additional mechanism of scattering. To this end,
we calculate the resistivity of a two-dimensional electron gas with two filled transverse subbands in a presence of 
electron--electron and impurity scattering. Though the collisions between the electrons do not directly affect 
the current in such systems, they cause a redistribution of the electrons between the Fermi contours, which results in a
noticeable change of resistivity for realistic mechanisms of impurity scattering.
\end{abstract}

\maketitle

\section{Introduction}

\nnn{In systems with a simple parabolic spectrum, electron--electron scattering does not contribute to the resistivity because
of their Galilean invariance. In other words, the electron--electron collisions do not change the current because it is proportional to the total momentum of electrons, which is conserved in such collisions. The situation is more interesting if 
the spectrum of electrons has several branches, which may break the Galilean invariance.}
The effects of electron--electron scattering on the transport properties of different conductors that lack Galilean 
invariance have been a subject of interest for many years. Typically, the systems under investigation included 
two types of charge carriers with different charges or effective masses. First of all, a number of papers \nnn{theoretically}
addressed scattering between electrons and holes in semimetals \cite{Baber37,Kukkonen76,Gantmakher78,Li18}. Another group of 
papers dealt \nnn{experimentally and theoretically} with mutual scattering of holes from spin-split subbands in GaAs 
heterostructures  \cite{Kravchenko99,Murzin98,Hwang03} in order to explain the unusual temperature of resistivity and 
magnetoresistance in them.  The cyclotron resonance in Si inversion layers in a presence of collisions between electrons in 
different valleys  was investigated in \cite{Appel78}. Several years ago, the electrical conductivity was calculated for non-
Galilean electron systems with several geometries of Fermi surface \cite{Maslov11,Pal12-LJP}. 
Very recently,  it was calculated for a two-dimensional electron gas with Rashba spin-orbit coupling
\cite{Nagaev20}. \nnn{Most theoretical papers predicted an increase of resistivity proportional to the square of temperature.
The experiment \cite{Murzin98} also revealed a decrease of the temperature derivative of resistivity at sufficiently high
temperatures.} All these papers have in common that  individual two-particle collisions do not conserve the current 
despite momentum conservation.  The reason is that the colliding particles have either opposite charges or different effective 
masses, so the current is not proportional to the total momentum of charge carriers.

In this paper, we show that the electron--electron collisions contribute to the resistivity even if the electrons are 
Galilean-invariant provided that there is an additional mechanism of scattering. To this end, we consider a two-dimensional 
(2D) electron gas formed in a semiconductor heterostructure with two populated subbands of transverse quantization. 
The Fermi surface of such systems is doubly connected, but the effective masses are equal in both subbands, hence any individual 
electron--electron collision cannot change the current. Nevertheless these collisions affect the resistivity if either the 
electron--impurity scattering rates on the two Fermi contours are different or there is a sufficiently strong intersubband impurity 
scattering. The reason is that the two-particle collisions redistribute nonequilibrium electrons between the two contours. As a 
result, the resistivity increases with temperature to a new finite value. We analyze different mechanisms of impurity scattering
and determine the optimal conditions for observing this effect.

Except Ref. \cite{Nagaev20}, all cited calculations of the electron--electron contribution to the resistivity used certain
simplifying assumptions. In particular, they neglected the scattering processes that involve only one type of charge carriers.
Here we take into account several possible scattering channels and show that the collisions involving only electrons at the same
Fermi contour are essential at intermediate temperatures.

The paper is organized as follows. Section \ref{sec:gen} presents the model, the kinetic equation, and general expressions for 
the electron--electron and electron--impurity collision integrals. In Section \ref{sec:sol}, the kinetic equation is solved
for arbitrary mechanism of impurity scattering. In Section \ref{sec:types}, the results for different types of impurities are compared, and Section \ref{sec:summary} contains the summary of the results. Some lengthy expressions are given in the Appendix.

\section{The model and kinetic equation} \label{sec:gen}

Consider a 2D semiconductor heterostructure with parabolic dispersion law and Fermi level crossing two subbands of transverse
quantization. In what follows, we will designate the lower and upper subbands by $n=1$ and $n=2$, respectively. Hence the 
dispersion laws of electrons in these subbands are $\eps_n(\bp) = p^2/2m + \delta_{n2}\,\Delta_0$, where $\Delta_0$ is the 
subband splitting.
At low temperatures, the resistivity of this system is dominated by electron--impurity and electron--electron scattering, so
the kinetic equation for the nonequilibrium electron distribution is of the form
\be
 -\frac{e{\bm E}{\bm v}_{n}}{T}\,\bar{f}\,(1-\bar{f}) = I_{n}^{imp} + I_{n}^{ee},
 \label{Boltz-1}
\ee
where $\bm E$ is the electric field, ${\bm v}_n=\partial\eps_n/\partial\bp$, and $\bar{f}(\eps_n)$ is the equilibrium Fermi function. The electron--impurity collision integral with account taken of intersubband scattering was obtained in many papers
\cite{Siggia70,Hai98,Inoue93,Gonzalez99} and in the Born approximation is given by the equation
\begin{multline}
 I_n^{imp}(\bp) = \frac{2\pi}{\hbar}\sum_{n'}\int\frac{d^2p'}{(2\pi\hbar)^2}\,
 \delta(\eps_n-\eps_{n'})
 \\ \times
 |U_{nn'}(\bp-\bp')|^2\,[f_{n'}(\bp') - f_n(\bp)],
 \label{Imp-1}
\end{multline}
where $U_{nn'}(\bp-\bp')$ is the matrix element of the impurity potential between the electron states $(\bp,n)$ and
$(\bp',n')$. The specific form of these matrix elements for different kinds of impurities will be discussed in Section  \ref{sec:types}.

The electron--electron collision integral may be written in the standard form as
\begin{multline}
 I^{ee}_{n}(\bp) = \sum_{n_1} \sum_{n_2} \sum_{n_3}
 \int\frac{d^2p_1}{(2\pi\hbar)^2} \int\frac{d^2p_2}{(2\pi\hbar)^2} \int d^2p_3\,
 \\ \times
 \delta(\bp + \bp_1  -  \bp_2 - \bp_3)\,
 \delta(\eps_{n} + \eps_{n_1} - \eps_{n_2} - \eps_{n_3})
 \\ \times
 W_{\lambdas}(\bp\bp_1, \bp_2 \bp_3)
 \\ \times
 \bigl[ (1-f)(1-f_1)\,f_2\,f_3 - f\,f_1\,(1-f_2)(1 - f_3) \bigr].
 \label{Iee-1}
\end{multline}
In the limit of weak electron--electron interaction, the scattering probabilities may be calculated in the Born
approximation. For simplicity, we assume that the interaction potential $V$ is short-ranged due to the presence of 
a nearby screening gate. If the distance to the gate $d_0$ is smaller than the Fermi wavelength,
\be
 W_{n..n_3} \equiv \frac{2\pi}{\hbar}\,\la n n_1|V|n_2 n_3\ra^2,
 \label{W}
\ee
where 
$
 \la n n_1|V|n_2 n_3\ra = 4\pi e_0^2\,d_0\,\kappa^{-1}\,\delta_{nn_2}\,\delta_{n_1 n_3},
$
$e_0$ is the electron charge, and $\kappa$ is the dielectric constant. It was taken into account here that for wave
vectors smaller than $1/d_0$, the longitudinal Fourier transform of $V$ weakly depends on transverse coordinates
and that the transverse wave functions for $n=1$ and $n=2$ are orthogonal.

As the system is rotationally symmetric in the plane of the electron gas, it is convenient to seek the solution of Eq.
\eqref{Boltz-1} in the form
\be
 f_{n}(\bp) = \bar{f}(\eps) + \bar{f}(\eps)\,[1 - \bar{f}(\eps)]\,
 C_{n}(\eps)\cos\p,
 \label{ansatz-f}
\ee
where the energy $\eps$ is measured from $E_F$ and $\p$ is the angle between $\bm E$ and $\bp$. With this substitution,
the electron--impurity collision integral is easily brought to the form
\be
 I_n^{imp}(\eps,\p) = (\Gamma_x\,C_{3-n} - \Gamma_n\,C_n)\,\cos\p\,\bar{f}\,(1-\bar{f}),
 \label{Imp-2}
\ee
where the intrasubband scattering rates 
\begin{multline}
 \Gamma_n = \frac{m}{2\pi\hbar^3} \int_{-\pi}^{\pi} d\chi\,
 \bigl[ |U_{12}(\bp-\bp')|^2 
 \\+ 
 |U_{nn}(\bp-\bp')|^2\,(1 - \cos\chi)\bigr]
 \label{Gn-1}
\end{multline}
are always positive, but the intersubband scattering rate 
\be 
 \Gamma_x = \frac{m}{2\pi\hbar^3} \int_{-\pi}^{\pi} d\chi\,|U_{12}(\bp-\bp')|^2\,\cos\chi
 \label{Gx-1}
\ee
may be in general of either sign. Note that $\Gamma_n > |\Gamma_x|$ regardless of the scattering potential.

The electron--electron collision integral is treated in the way similar to Ref. \cite{Nagaev20}. A
substitution of Eq. \eqref{ansatz-f} into Eq. \eqref{Iee-1} brings it to the form
\begin{multline}
 I_{n}^{ee}(\eps,\p) = m \sum_{n_1} \sum_{n_2} \sum_{n_3} W_{\lambdas}
 \int d\eps_1 \int d\eps_2 \int d\eps_3
 \\ \times
 \delta(\eps + \eps_1 - \eps_2 - \eps_3)\,
 (1 - \bar{f})(1 - \bar{f}_1)\,\bar{f}_2\,\bar{f}_3
 \\ \times
  \int d\p_1\int\frac{d^2p_2}{(2\pi\hbar)^2} \int\frac{d^2p_3}{(2\pi\hbar)^2}\,
 \delta(\eps_{n_2} - \eps_2)\,\delta(\eps_{n_3} - \eps_3)
 \\ \times
 \delta(\bp + \bp_1 - \bp_2 - \bp_3) 
 \bigl[C_{n_2}(\eps_2)\cos\p_2 + C_{n_3}(\eps_3)\cos\p_3 
 \\- 
 C_{n}(\eps)\cos\p - C_{n_1}(\eps_1)\cos\p_1 \bigr].
 \label{Iee-2}
\end{multline}
Furthermore, the cosines in the last factor of the integrand  may be expressed in terms of $p\ldots p_3$, $\p$, 
and $\p_1$. Assume that all the quantities except the distribution functions $\bar f$ and $C_n$ are energy-independent 
near the Fermi level. Subsequently integrating in Eq. \eqref{Iee-2} over $\bp_3$, $\bp_2$, and $d\p_1$, one obtains
\begin{multline}
 I_n^{ee} = \cos\p\,\frac{\Gamma_{ee}}{T^2} \int d\eps'\,K(\eps,\eps')
 \biggl\{ 
  Q_n\, \left[C_n(\eps') - C_n(\eps)\right]
  \\ +
  \Psi_n\,\frac{p_n C_{3-n}(\eps') - p_{3-n} C_n(\eps')}{p_1+p_2} 
 \biggr\},
 \label{Iee-3}
\end{multline}%
where the effective electron--electron scattering rate is given by
$\Gamma_{ee} = {8\pi\,e_0^4\,d_0^2\,m^3 T^2}/{\kappa^2\hbar^5 p_1 p_2}$
and
\be
 K(\eps,\eps') = [1 - \bar{f}(\eps)]\,\frac{\eps-\eps'}{e^{(\eps-\eps')/T} - 1}\,\bar{f}(\eps').
 \label{K}
\ee
It is assumed that $C_n$ are even functions of $\eps$.

The first term in Eq. \eqref{Iee-3} is similar to the expression for the
2D conductors with singly connected Fermi surface . The coefficients
\be
 Q_n = 2\,\frac{p_n + 2\,p_{3-n}}{p_n}\,\ln\!\left(\frac{p_n^2}{2mT}\right)
 \label{Q}
\ee
exhibit a logarithmic singularity at $T\to 0$ that results from head-on or small-angle collisions \cite{Hodges71,Giuliani82}. 
However this term vanishes 
identically if $C_n(\eps) = {\rm const}$, and for this reason it does not contribute to the electric resistivity if taken alone.
Note that it originates not only from the collisions in which all the initial and final states belong to the same Fermi contour,
but also from the scattering processes in which both the initial and the final states are at different contours.

The second term in Eq. \eqref{Iee-3} is nonzero even for energy-independent $C_n$ and arises only from collisions of electrons 
with both initial and final states at different Fermi contours. The quantities $\Psi_n$ are given by the equations
\be
 \Psi_n = \frac{p_1+p_2}{p_n}\,\ln\!\left(\frac{p_1+p_2}{p_1-p_2}\right).
 \label{Psi}
\ee
The structure of this term emanates from the fact that the collision integral must be turned into zero by a distribution
$f_n(\bp) = \bar{f}(\eps_n) + {\bm u}\bp\,\bar{f}(\eps_n)\,[1 - \bar{f}(\eps_n)]$,
i.~e. with $p_1 C_2 = p_2 C_1$ \cite{Nagaev21}. Another condition $p_1\Psi_1 = p_2\Psi_2$ stems from the Galilean invariance
of the system and the current conservation by the electron--electron collisions. The second term in Eq. \eqref{Iee-3} is
of crucial importance for the effects considered here.

\section{Electrical conductivity} \label{sec:sol}

Equation \eqref{Boltz-1} results in a system of two integral equations in $C_n(\eps)$. To solve it, one may use  the
method proposed by Brooker and Sykes \cite{Brooker68} for calculating thermal conductivity of Fermi liquid. Upon a 
substitution of Eq. \eqref{ansatz-f} into Eq. \eqref{Boltz-1}, reduce both parts by $\cos\p\,\sqrt{\bar{f}\,(1-\bar{f})}$ and introduce new variable
\be
 \rho_{n}(\eps) = \sqrt{\bar{f}\,(1-\bar{f})}\,C_{n}(\eps).
 \label{ansatz-C}
\ee
As a result, the kernel $K(\eps,\eps')$ of the integral in Eq. \eqref{Iee-3} is replaced by a function of $\eps'-\eps$, and
the integral equation may be brought to the differential form by a Fourier transform in $\eps$ with the parameter $u$. A 
subsequent replacement of the independent variable $\xi= \tanh(\pi Tu)$ brings these equations to the form
\begin{multline}
 \Gamma_{ee} Q_{n}\,(\hat{L} + 2)\,\rho_n(\xi)
 +
 2\,\Gamma_{ee} \Psi_n\,\frac{p_n\,\rho_{3-n} - p_{3-n}\,\rho_n}
                                {p_1 + p_2}
\\ -
 \frac{1}{\pi^2}\,\frac{\Gamma_n\,\rho_n - \Gamma_x\,\rho_{3-n}}{1-\xi^2}\,
 =
 -\frac{eEv_n}{\pi \sqrt{1 - \xi^2}},     
 \label{Boltz-xi}                            
\end{multline}
where $\hat L$ is the differential operator
\be
 \hat{L}\,\psi = \frac{\partial}{\partial\xi}\biggl[(1-\xi^2)\,\frac{\partial\psi}{\partial\xi}\biggr]
 - \frac{\psi}{1-\xi^2}.
 \label{L}
\ee
As the solutions of Eqs. \eqref{Boltz-xi} are even functions of $\xi$, they may be presented in the form of a series
\be
 \rho_{n}(\xi) = \sum_{r=0}^{\infty} \gamma_{nr}\,\psi_{2r}(\xi),
 \label{series}
\ee
where $\psi_r(\xi)$ are the eigenfunctions of operator $\hat L$ with eigenvalues $-(r+1)(r+2)$ \cite{Landau-book}.
A substitution of these expansions into Eqs. \eqref{Boltz-xi} results in an infinite system of equations for the
coefficients $\gamma_{nr}$ of the form
\begin{multline}
 2\,\Gamma_{ee} \biggl[
  r\,(2r+3)\,Q_n\,\gamma_{nr} 
  + 
  \Psi_n\,\frac{p_{3-n}\,\gamma_{nr} - p_n\,\gamma_{3-n,r}}{p_1 + p_2}
 \Biggr]
 \\+
 \frac{1}{\pi^2} \sum_{r=0}^{\infty} Y_{rs}\,(\Gamma_n\,\gamma_{ns} - \Gamma_x\,\gamma_{3-n,s})
 =
 \frac{eEv_n}{\pi}\,X_r,
 \label{Boltz-inf}
\end{multline}
where $Y_{rs}$ are the matrix elements of $(1-\xi^2)^{-1}$ between $\psi_{2r}$ and $\psi_{2s}$, and $X_r$ are
the projections of $(1-\xi^2)^{-1/2}$ on $\psi_{2r}$. The explicit expressions for these quantities are given in
the Appendix. Once the coefficients $\gamma_{nr}$ are found, the current density is calculated as the sum
\begin{multline}
 j = \frac{e}{2\pi\hbar^2} \sum_n \int d\eps\,\bar{f}\,(1 - \bar{f})\,p_n\,C_n(\eps)
\\=
 \frac{e}{4\pi^2\hbar^2} \sum_n p_n \sum_r X_r\,\gamma_{nr}.
 \label{j-1}
\end{multline}
In general, the system \eqref{Boltz-inf} may be solved only numerically, but an analytical solution is possible
in the limiting cases of vanishing or very strong electron-electron scattering. If $\Gamma_{ee}=0$, Eq. \eqref{Boltz-1}
with the electron--impurity collision integral Eq. \eqref{Imp-2} is easily solved in $C_n$, and the first part of Eq.
\eqref{j-1} readily gives
\be
 \sigma_0 = \frac{e^2}{2\pi\hbar^2}\,\frac{\Gamma_1\, p_2 v_2 + \Gamma_2\, p_1 v_1 + \Gamma_x\,(p_1 v_2 + p_2 v_1)}%
 {\Gamma_1 \Gamma_2 - \Gamma_x^2}
 \label{sigma-0}
\ee
in agreement with previous results for two-subband conductors. In the opposite limit of a very strong electron--electron 
scattering, it is sufficient to keep in the system \eqref{Boltz-inf} only the pair of equations with $r=0$. However one
cannot just neglect the electron--impurity scattering because the system would be satisfied by any solution with
$p_2\gamma_{10}=p_1\gamma_{20}$ and hence be degenerate. To lift the degeneracy, one has to keep the electron--impurity
scattering rates nonzero and isolate the most singular contribution in them to $\gamma_{n0}$. The resulting
coefficients are inversely proportional to a linear combination of the impurity-scattering rates  and satisfy the above 
ratio, so that \nnn{the currents carried by the subbands are proportional to the electron densities in them}.
Making use of the
explicit expressions for $\Psi_n$ Eq. \eqref{Psi} and the expression for $j$ Eq. \eqref{j-1}, one easily obtains the 
corresponding conductivity  
\be
 \sigma_{\infty} = \frac{e^2}{2\pi\hbar^2}\, \frac{(p_1v_1 + p_2v_2)^2}
 {\Gamma_1\,p_1  v_1  + \Gamma_2\,p_2 v_2  - \Gamma_x\,(p_1 v_2 + p_2 v_1)}.
 \label{sigma-inf}
\ee
The $\sigma_0(E_F)$ and $\sigma_{\infty}(E_F)$ curves 
vary in shape depending on the ratios of $\Gamma_n$ and $\Gamma_x$,  but 
the difference of these conductivities
\be
 {\sigma_0 - \sigma_{\infty}}
 \propto
 {[(\Gamma_1 - \Gamma_2)\,p_1\,p_2 + \Gamma_x\,(p_1^2 - p_2^2)]^2}
 \label{difference}
\ee
is positive regardless of impurity type. The quadratic dependence of the conductivity difference Eq. \eqref{difference}
on the difference of the scattering rates is the consequence of the structure of the second term of the collision integral
Eq. \eqref{Iee-3} and the general relation between $\Psi_1$ and $\Psi_2$, so it does not depend on the particular choice
of the interaction potential $V$. 

\begin{figure}
 \includegraphics[width=1.0\columnwidth]{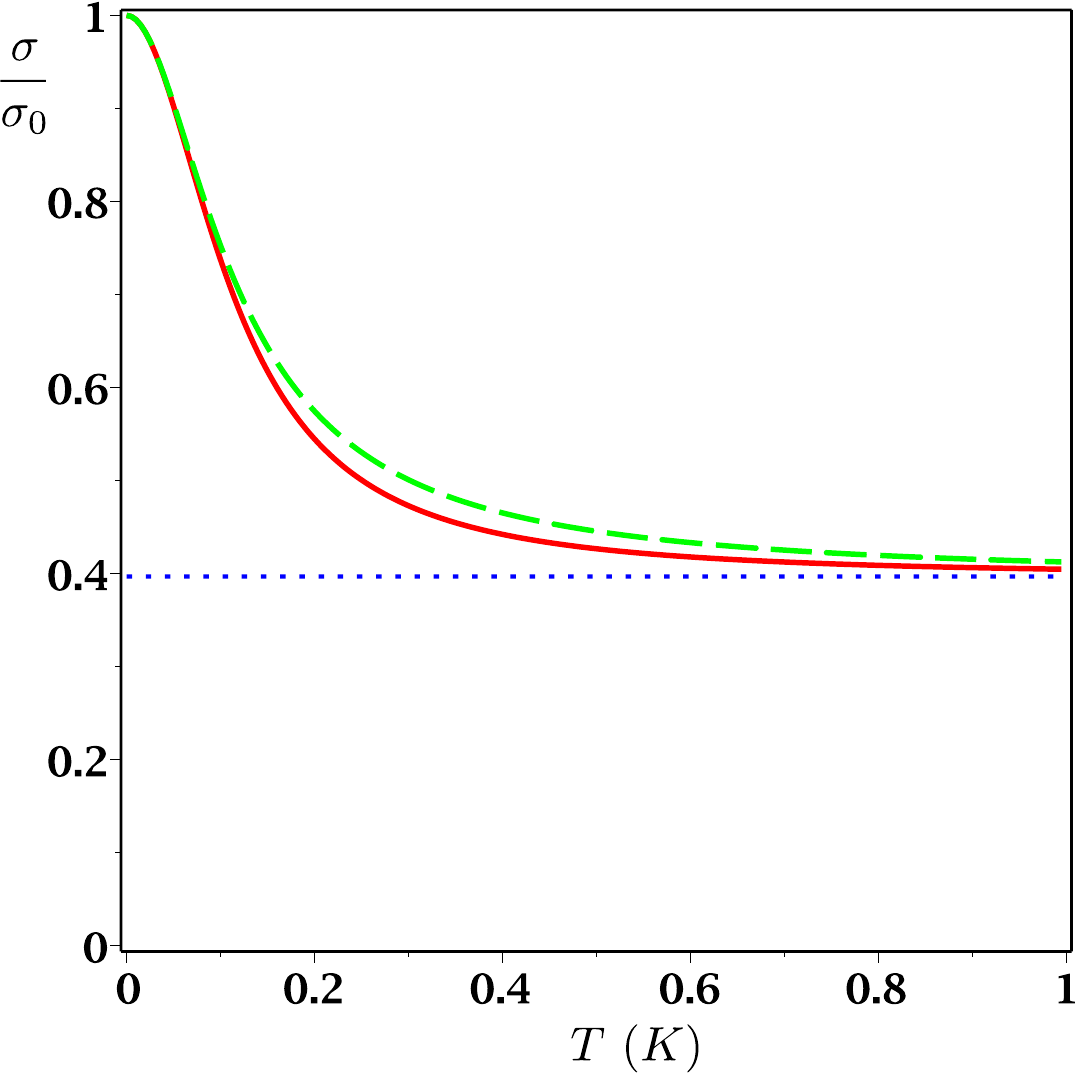}
 \caption{\label{fig:vs_T} The $\sigma(T)/\sigma_0$ dependence for \nnn{a GaAs heterostructure with $\Delta_0=10$ meV, 
 $E_F=15$ meV, $d_0=20$ nm, $\Gamma_1=2.5\times 10^{11}$ s$^{-1}$, $\Gamma_2=0.1\,\Gamma_1$, and $\Gamma_x=0$.} The dashed 
 line shows the same quantity calculated with $Q_n=0$.}
\end{figure}
For arbitrary temperatures and arbitrary relation between the electron--electron and impurity scattering, Eqs. 
\eqref{Boltz-inf} may be solved only numerically. The $\sigma(T)$ curve for \nnn{a GaAs heterostructure with $\Delta_0=10$ meV, 
$E_F=15$ meV, $d_0=20$ nm, $\Gamma_1=2.5\times 10^{11}$ s$^{-1}$, $\Gamma_2=0.1\,\Gamma_1$, and $\Gamma_x=0$} is shown in 
Fig.~\ref{fig:vs_T}. The dashed line shows the same curve calculated for $Q_n=0$,
i.~e. in the absence of the first term in the collision integral Eq. \eqref{Iee-3}. Though the solid and the dashed 
curves almost coincide in the low-$T$ and high-$T$ limits, they are noticeably different at intermediate temperatures.
This suggests that the intrasubband electron--electron scattering also contributes to the electric resistivity, in contrast
to the claim made in many previous papers \cite{Kukkonen76,Li18,Kravchenko99,Hwang03,Appel78,Maslov11,Pal12-LJP}. 
The difference between the two calculated conductivities does not appear to be parametrically small.

\section{Estimates for different types of impurities}   \label{sec:types}

\begin{figure}
 \includegraphics[width=1.0\columnwidth]{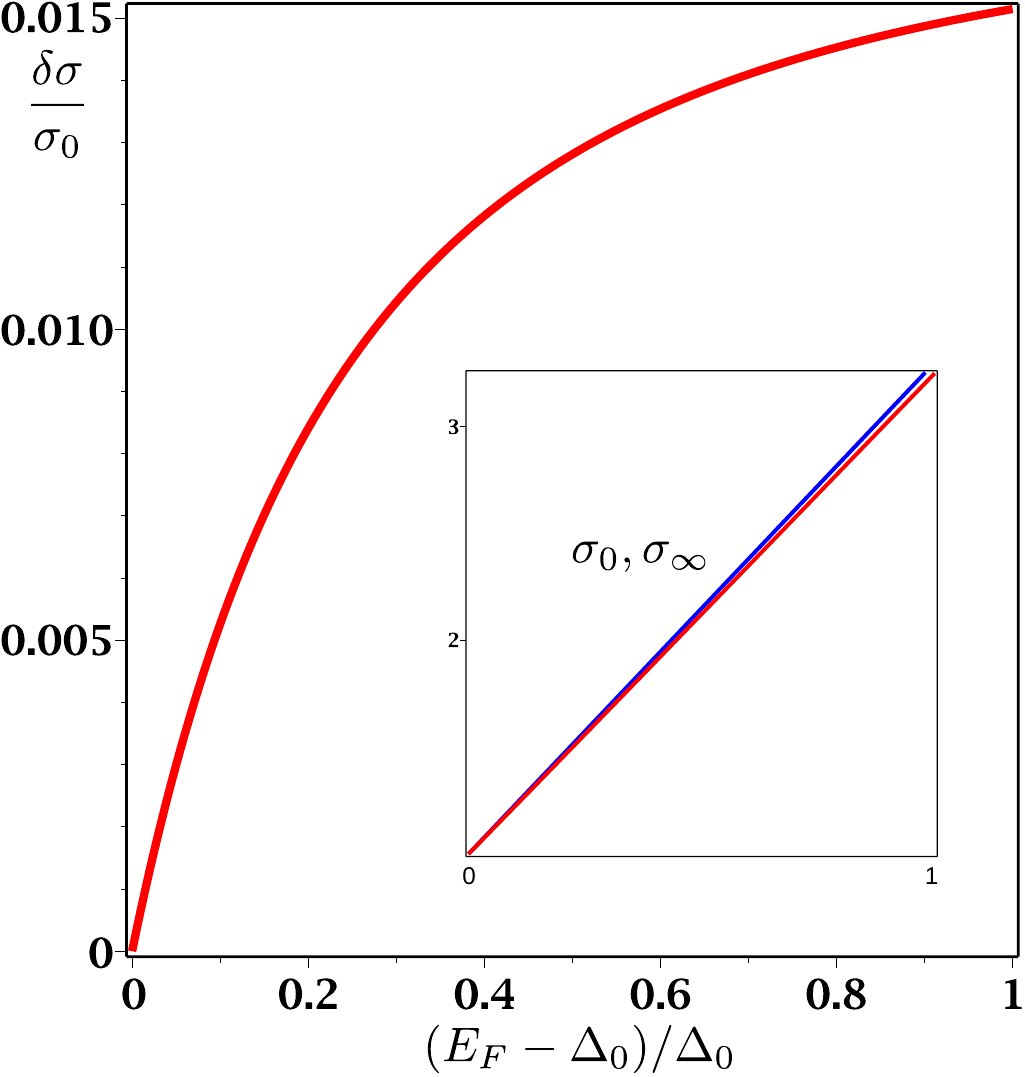}
 \caption{\label{fig:point} The dependence of the relative difference 
 $\Delta\sigma/\sigma_0=(\sigma_0 - \sigma_{\infty})/\sigma_0$ on the position
 of the Fermi level $E_F$ for point-like impurities. Inset separately shows $\sigma_0(E_F)$ (blue curve) and 
 $\sigma_{\infty}(E_F)$ (red curve) normalized to $\sigma_0(\Delta_0)$. }
\end{figure}

Estimate now the effect of electron--electron scattering for different types of impurities. The intersubband scattering
rate $\Gamma_x$ is typically much smaller than the intrasubband rates $\Gamma_n$, and therefore to maximize the difference
$\sigma_0-\sigma_{\infty}$, one has to find the type of impurities with maximum difference $|\Gamma_1-\Gamma_2|$.

First of all, consider neutral point-like impurities, which may be formed by atomic vacancies in the semiconductor.
In this case, the scattering potential may be written in the form
\be
 U(\br, z) = \Omega_0 \sum_i \delta(\br-\br_i)\,\delta(z-z_i),
 \label{point-1}
\ee
where $i$ labels impurities, $\br$ stands for the in-plane coordinates and $z$ is the transverse coordinate.
Therefore the impurity-averaged square of the matrix element is
\be
 |U_{nn'}(\bp-\bp')|^2 = \Omega_0^2\,n_{imp} \int dz\,\phi_n^2(z)\,\phi_{n'}^2(z),
 \label{point-2}
\ee
where $n_{imp}$ is the three-dimensional concentration of impurities and $\phi_n(z)$ are the transverse 
components of the electron wave functions corresponding to the subband $n$. It is immediately seen that $\Gamma_x=0$
in this case. The estimates made using $\phi_n$ for a triangular transverse confining potential give
$\Gamma_1/\Gamma_2 \approx 1.3$, which results only in a slight decrease of $\sigma$ with temperature about 1\% 
(Fig. \ref{fig:point}).

\begin{figure}
 \includegraphics[width=1.0\columnwidth]{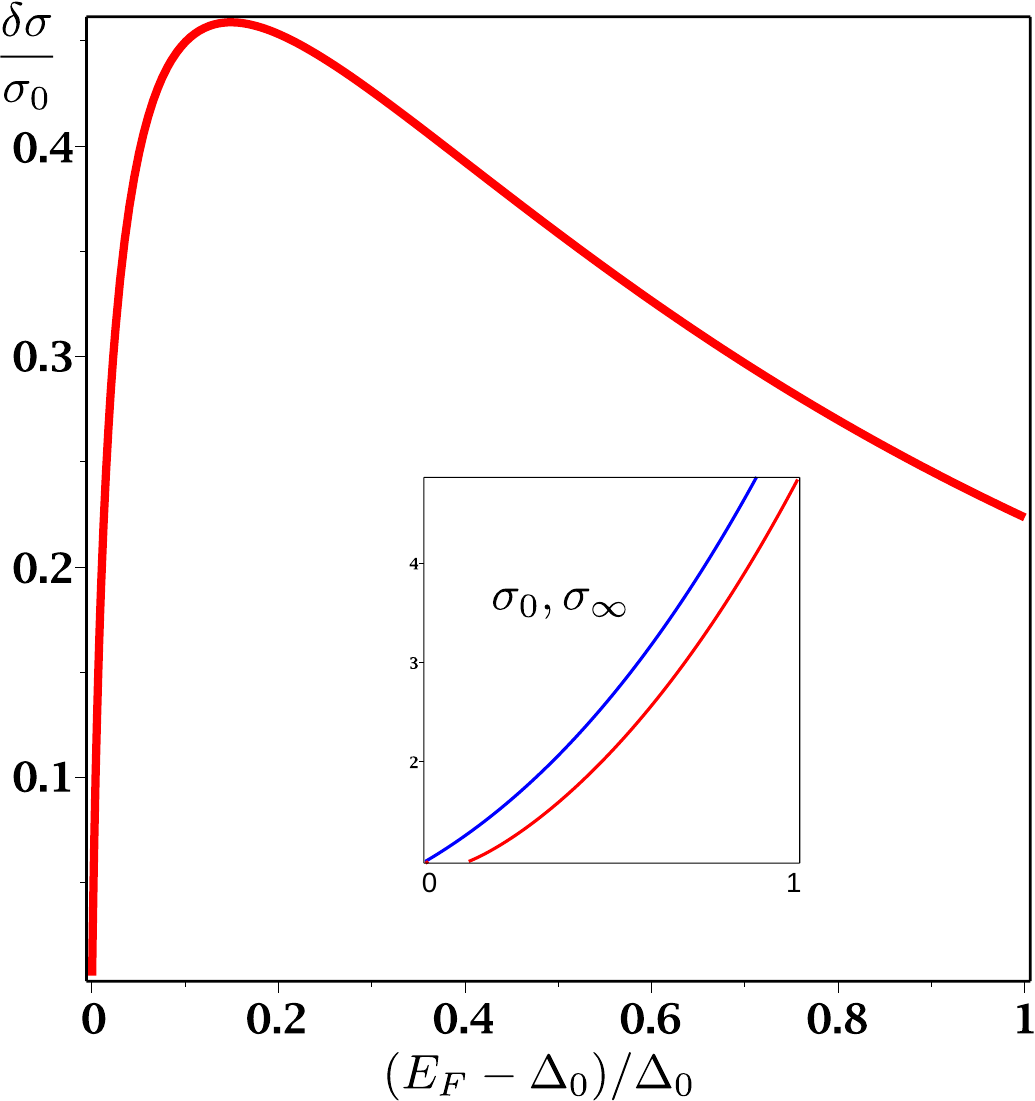}
 \caption{\label{fig:delta} The dependence of the relative difference 
 $\delta\sigma/\sigma_0=(\sigma_0 - \sigma_{\infty})/\sigma_0$ on the position
 of the Fermi level $E_F$ for delta-doped Coulomb impurities. The distance between the electron gas and the impurity
 layer is $h_0 = 3\,\hbar/p_1(\Delta_0)$. Inset  shows $\sigma_0(E_F)$ (blue curve) and 
 $\sigma_{\infty}(E_F)$ (red curve) normalized to $\sigma_0(\Delta_0)$. }
\end{figure}

Consider now a delta-doped semiconductor heterostructure where the 2D electron gas is located between the impurities and
the metallic gate
in such a way that the impurity--2D gas distance $h_0$ is much larger than the 2D gas--gate separation $d_0$ \cite{Ando82}.
The 2D Fourier transform of the unscreened potential of a single impurity 
$U_0(q) = 2\pi e_0^2\,\exp(-qh_0)/q$
is well known. Therefore in a presence of the gate, it becomes
\begin{multline}
 \bar{U}_0(q) = {2\pi e_0^2}\,q^{-1} \bigl\{ \exp(-qh_0) - \exp[-q\,(h_0+2d_0)] \bigr\}
\\
 \approx 4\pi e_0^2\,d_0\,\exp(-qh_0)
 \label{_U0}
\end{multline}
provided that $qd_0 \ll 1$. As  $\bar{U}_0(q)$ is independent of $z$, the matrix elements
$U_{12}$ and $\Gamma_x$ are zero in this approximation. Following Ref. \cite{DasSarma13}, one may bring Eq. \eqref{Gn-1}
for the intrasubband scattering rates to the form
\be
 \Gamma_n = 64\pi\,\frac{mN_{imp} e_0^4\, d_0^2}{\hbar^3}
 \int_0^1 d\xi\,\frac{\xi^2}{\sqrt{1-\xi^2}}\,
 e^{-4p_n h_0 \xi/\hbar},
 \label{Gn-2}
\ee
where $N_{imp}$ is the 2D concentration of impurities. If $p_nh_0 \ll \hbar$, the integral over $\xi$ equals $\pi/4$. In 
the opposite limit, it scales as $(p_n h_0/\hbar)^{-3}$. If the Fermi level crosses the upper subband near its bottom and
$p_2 \ll p_1$, it is quite possible that $p_1 h_0 >\hbar$ while $p_2 h_0<\hbar$ and $p_{1} d_0/\hbar <1$. 
In this case, $\Gamma_1 \ll \Gamma_2$ and the suppression of conductivity by electron--electron collisions may be about 
100\%  if $E_F-\Delta_0 < \Delta_0$ (Fig. \ref{fig:delta}).

\begin{figure}
 \includegraphics[width=1.0\columnwidth]{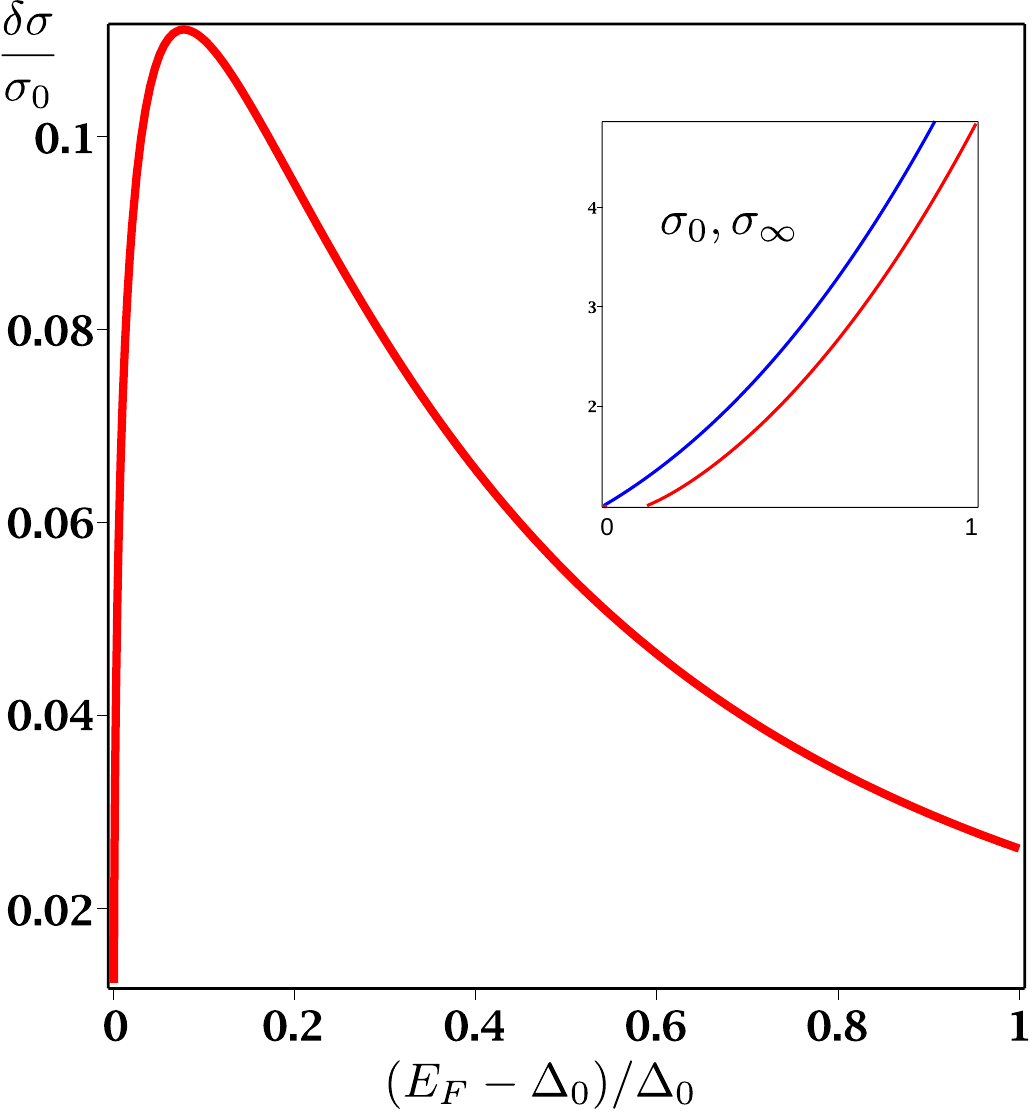}
 \caption{\label{fig:uniform} The dependence of the relative difference 
 $\delta\sigma/\sigma_0=(\sigma_0 - \sigma_{\infty})/\sigma_0$ on the position
 of the Fermi level $E_F$ for uniformly distributed Coulomb impurities. Inset  shows $\sigma_0(E_F)$ (blue curve) and 
 $\sigma_{\infty}(E_F)$ (red curve) normalized to $\sigma_0(\Delta_0)$. }
\end{figure}

Finally, if the charged impurities with three-dimensional concentration $n_{imp}$ are uniformly distributed over the 
volume of the semiconductor, the intrasubband scattering rates may be obtained by integrating Eq. \eqref{Gn-2} over $h_0$,
which readily gives
\be
 \Gamma_n = 16\pi\,{m n_{imp} e_0^4\,d_0^2}/{\hbar^2 p_n}.
 \label{Gn-3}
\ee
Therefore $\Gamma_1/\Gamma_2 = p_2/p_1$, and the maximum difference of 
$\sigma_0$ and $\sigma_{\infty}$
is about 10\% 
at $E_F \approx 1.1\,\Delta_0$ (Fig. \ref{fig:uniform}).

\begin{figure}
 \includegraphics[width=1.0\columnwidth]{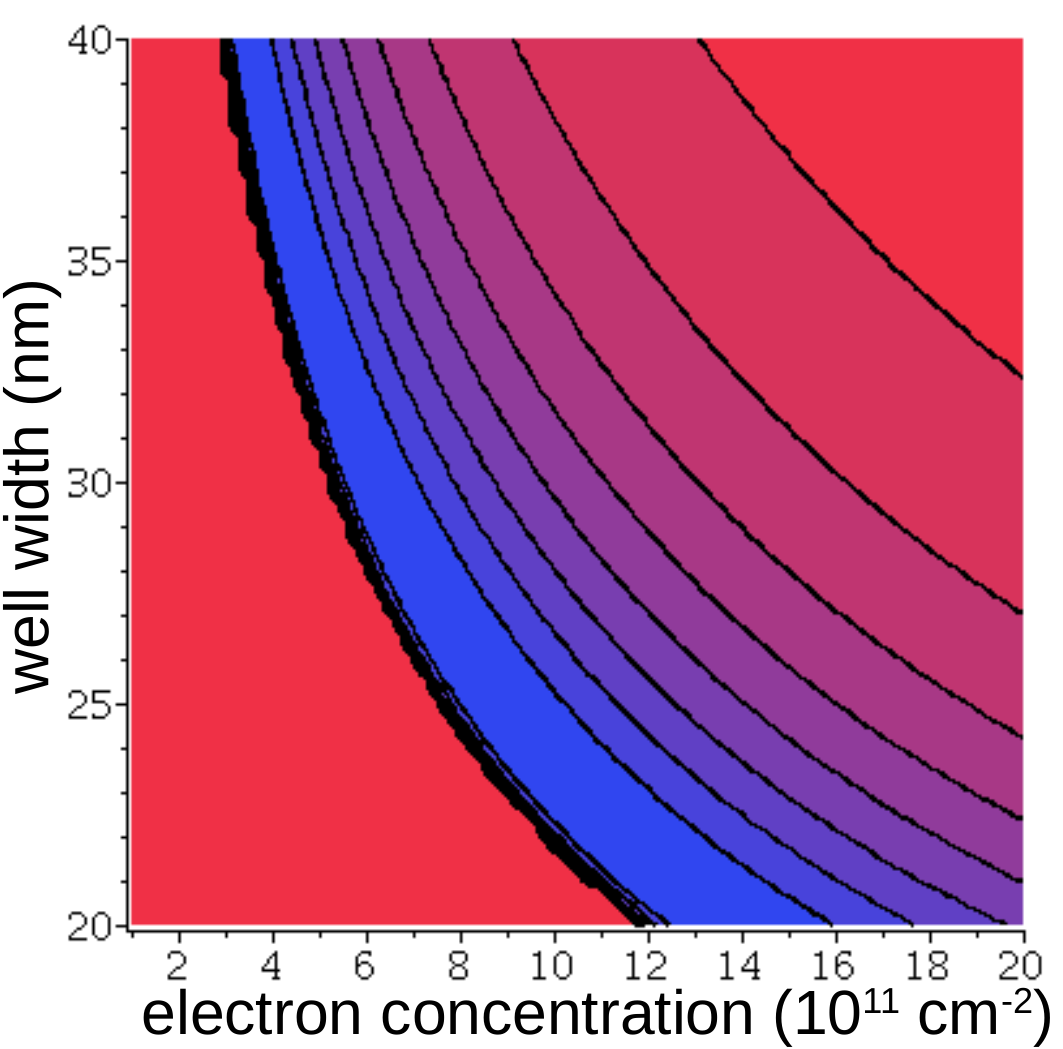}
 \caption{\label{fig:contour} Contour plot of $\delta\sigma/\sigma_0$ as a function of electron concentration and the width 
 of quantum well for the case of uniformly distributed Coulomb impurities. Blue color correspond to higher values of 
 $\delta\sigma/\sigma_0$, and red color to its smaller values. }
\end{figure}

A good candidate for observing the suppression of conductivity are AlGaAs/GaAs heterostructures. Currently,
these systems reached such a quality that the electron--electron scattering  is much stronger than the impurity
scattering at temperatures as low as 2 K \cite{Melnikov12,Gusev18}. Transport measurements in GaAs quantum wells
with two populated subbands are also possible \cite{Goran09,Dietrich12}. Figure \ref{fig:contour} shows the relative
correction to the conductivity calculated as a function of electron concentration and the width of a rectangular GaAs 
quantum well, which was used in experiments \cite{Goran09,Dietrich12}. In this approximation, the maximum value of 
$\delta\sigma/\sigma_0$ does not depend on the width of the well
and corresponds to the electron concentration slightly above the threshold for the population of the second subband. Wider
wells require lower electron concentrations for observing the effect, but for narrow wells the range of concentrations where 
the effect exists is wider. The suppression of conductivity by electron--electron scattering could be also observed 
in double quantum wells, where the overlap between the wave functions in them results in a formation of symmetric and 
antisymmetric states with a small energy separation, so the population of two subbands may be achieved at lower 
electron densities \cite{Gusev21}.

The above effects may be expected to take place at the 
temperatures about several Kelvins or lower, so the temperature-dependent correction to the resistivity from 
electron--electron scattering should saturate long before the electron--phonon scattering comes into play. 
Note that this effect
is purely semiclassical and therefore it is much larger than the correction the the conductivity from quantum interference
between electron-electron interaction and impurity scattering \cite{Altshuler85}. For two-dimensional systems, this correction 
is of the order of $e_0^2/\hbar \sim 10^{-4}\,\Omega^{-1}$, whereas for high-quality  GaAs-AlGaAs heterosructures, the 
conductivity is on the order of $10^{-1}\,\Omega^{-1}$ and the semiclassical correction may be on the order of this conductivity.

\section{Conclusions} \label{sec:summary}

In conclusion, it was shown that the electron--electron collisions may contribute to the resistivity of Galilean-invariant
systems in a presence of other mechanisms of scattering. In a 2D electron gas with a parabolic spectrum and two populated 
subbands, it results in a significant increase of resistivity with temperature provided that the elastic scattering 
rates in these subbands are essentially different. The reason is that these collisions redistribute nonequilibrium electrons 
that carry the current between the subbands in favor of the subband with stronger elastic scattering. The effect is maximal if 
the elastic scattering is caused by delta-doped Coulomb impurities located from the 2D electrons further than the screening 
length and the Fermi level is not too high above the bottom of the upper subband.

\nnn{As the expressions for $\sigma_0$ and $\sigma_{\infty}$ do not contain any parameters of electron--electron scattering,
the comparison of the low-temperature and high-temperature conductivities  allows a separate determination of the 
impurity-scattering rates of electrons $\Gamma_1$ and $\Gamma_2$ in both subbands.}

It was also shown that the standard assumption that the intraband electron--electron scattering does not affect the
resistivity of multiband systems holds only in the limit of very frequent two-particle collisions. Actually,
this scattering is essential at intermediate temperatures where the energy dependence of nonequilibrium electron 
distribution deviates from the simple derivative of the Fermi function.

Though the analytical expressions and numerical results were obtained for the particular case of the Coulomb interaction
of electrons screened by a metallic gate, the general conclusions of this paper remain valid for arbitrary interactions
because they stem only from the Fermi statistics and conservation laws.

\begin{acknowledgements}

I am grateful to Vadim Khrapai for a discussion.

\end{acknowledgements}

\appendix*
\section{Explicit expressions for some quantities}

The normalized eigenfunctions of differential operator $\hat L$ Eq. \eqref{L} are given by 
the expressions
\be
 \psi_r(\xi) = \sqrt{\frac{(2r+3)(r+2)}{8\,(r+1)}} \sqrt{1-\xi^2}\,P_r^{(1,1)}(\xi),
 \label{psi_r}
\ee
where $P_r^{(1,1)}(\xi)$ are Jacobi polynomials. The coefficients of expansion of $(1-\xi^2)^{-1/2}$
in these functions are given by 
\be
 X_r = \int_{-1}^1 d\xi\, \frac{\phi_{2r}(\xi)}{\sqrt{1-\xi^2}}
 =
 \sqrt{\frac{4r+3}{(2r+1)(r+1)}}.
 \label{X}
\ee
The matrix elements of $1/(1-\xi^2)$ between the eigenfunctions of $\hat L$ are given by the equation
\begin{multline}
 Y_{rs} = \int_{-1}^1 d\xi\,\frac{\phi_{2r}(\xi)\,\phi_{2s}(\xi)}{1 - \xi^2}
 =
 \frac{\min(r,s) + 1/2}{\max(r,s) + 1}
\\ \times
 \sqrt{\frac{ (4r+3)(r+1)(4s+3)(s+1) }{ (2r+1)(2s+1) }}.
 \label{Y}
\end{multline}

\bibliography{books,ee,imp2,2sub-exp}

\end{document}